\documentstyle[12pt]{article}

\begin{document}
\begin{titlepage}
\begin{center}
{\Large\bf Can the polarization of the strange quarks
\vskip 0.5cm
in the proton be positive~? }
\end{center}
\vskip 2cm
\begin{center}
{\bf Elliot Leader}\\
{\it Imperial College, London, UK}\\
\vskip 0.5cm {\bf Dimiter B. Stamenov\\
{\it Institute for Nuclear Research and Nuclear Energy\\
Bulgarian Academy of Sciences, Sofia, Bulgaria }}\\
\end{center}

\vskip 0.3cm
\begin{abstract}
Recently, the HERMES Collaboration at DESY, using a leading order QCD
analysis of their data on semi-inclusive deep inelastic production of charged
hadrons, reported a marginally positive polarization for the strange
quarks in the proton. We argue that a non-negative polarization is almost
impossible.\\


\end{abstract}
\end{titlepage}
\newpage
\setcounter{page}{1}

There is, at present, a major experimental drive (HERMES at DESY, COMPASS at
CERN) to determine the polarized sea-quark densities $\Delta \bar u(x, Q^2),
~\Delta \bar d(x, Q^2),~ \Delta s(x, Q^2)$ and $\Delta \bar s(x, Q^2)$,
as well as the polarized gluon
density $\Delta G(x, Q^2)$. These are being studied using polarized
semi-inclusive deep inelastic (SIDIS) reactions of the type
$l + p \rightarrow l + h + X$ where $h$ is an identified hadron and the
initial lepton and proton are longitudinally polarized.

Recently the HERMES group has presented preliminary data on the
polarized strange quark sea \cite{HERMES}, suggesting, in a
leading order QCD analysis, that ($\Delta s + \Delta \bar s)(x)$
at $Q^2=2.5~GeV^2$ is marginally positive, whereas in all analyses
of {\it inclusive} DIS \cite{inclDIS}, it is found that $(\Delta
s + \Delta \bar s)(x, Q^2)$ is significantly negative. We shall
argue in this note that a positive strange quark polarization is
almost impossible.

It has to be understood that there is a key difference between the
determination of the {\it non-strange} polarized sea-quark densities
$(\Delta \bar u, ~\Delta \bar d)$ and the strange sea contribution
$(\Delta s + \Delta \bar s)(x, Q^2)$. In inclusive DIS one can, in principle,
only determine combinations like $\Delta q + \Delta \bar q$. This implies
that even with perfect, error-free data we would know absolutely nothing
about $\Delta \bar u$ and $\Delta \bar d$ (note that in papers where these
densities are presented additional assumptions like SU(3) symmetric sea, etc.
have been used). But quite the opposite holds for
$(\Delta s+\Delta \bar s)(x, Q^2)$. It is completely determined subject, of
course to errors, in inclusive DIS experiments. In all of the many
independent analyses it turns out that the first moment

\begin{equation}
\delta s(Q^2) \equiv \int _{0} ^{1} dx [\Delta s(x,Q^2)+
\Delta \bar s(x,Q^2)]
\label{1}
\end{equation}
is significantly {\it negative}.

Consider the first moment $\Gamma_{1}^{p}(Q^2)$ of the measured
spin-dependent structure function $g_{1}^{p}(x,Q^{2})$. One has
in leading order QCD (more correctly, in leading logarithmic
approximation LLA),

\begin{equation}
 \Gamma_{1}^{p}(Q^2) \equiv \int_{0}^{1}dx g_{1}^{p}(x,Q^{2})
 =\frac{1}{6} [~\frac{1}{2} a_{3} + \frac{5}{6} a_{8} + 2\delta s(Q^2)~]
\label{2}
\end{equation}
where $a_{3}$ and $a_{8}$ are hadronic matrix elements of the
third and eighth components of the Cabibbo octet of axial-vector
currents which control the $\beta$-decays of the neutron
($a_{3}$) and the hyperons ($a_{8}$).

Now $a_{3}$ is known to high precision: $a_{3}=g_A=1.2670\pm
0.0035$ \cite{PDG}, and this determination relies only upon the
assumption of isotopic spin independence of the strong
interactions. On the other hand, the value usually attributed to
$a_{8}$, namely $a_{8}=3F-D$, is a consequence of the $SU(3)_f$
flavour symmetry treatment of the hyperon $\beta$-decays. Its
value (see the second ref. in \cite{inclDIS}) obtained on the
basis of updated $\beta$-decay constants is
\begin{equation}
a_{8}=3F-D=0.585\pm 0.025~.
\label{3FD}
\end{equation}
\def\thefootnote{\dagger}
While isospin symmetry is not in doubt, there is some question
about the accuracy of assuming $SU(3)_f$ symmetry in analyzing
hyperon $\beta$-decays. According to Ratcliffe \cite{Ratcliffe}
symmetry breaking effects are small, of order of 10\%. The recent
KTeV experiment at Fermilab \cite{KTeV} supports this assessment.
Their results of the $\beta$-decay of $~\Xi^0~, \Xi^0 \rightarrow
\Sigma^{+}e\bar{\nu}$, are all consistent with exact $SU(3)_f$
symmetry. Taking into account the experimental uncertainties one
finds that $SU(3)_f$ breaking is at most of order $20\%$. We
therefore conclude that it is almost impossible that $a_{8}$ lies
outside the range\footnote{Note that more extreme values of $a_8$
have emerged in some symmetry breaking models which study not
just octet hyperon $\beta$-decays, but also baryon magnetic
moments \cite{Gerasimov} and baryon decuplet $\beta$-decays
\cite{Manohar}. However, the predictions of these models for the
$~\Xi^0\rightarrow \Sigma^{+}~$ $\beta$-decay do not agree with
the experimental results of KTeV Collaboration. In addition, it
is the hyperon $\beta$-decays which are most relevant for the
matrix element $a_{8}$ needed in polarized DIS.}
\begin{equation}
0.47\leq a_{8} \leq 0.70.
\label{rangea8}
\end{equation}

Let us now return to Eq. (\ref{2}) and rewrite it in the form
\begin{equation}
a_{8}=\frac{6}{5}[~ 6\Gamma_{1}^{p}(Q^2)-\frac{1}{2}a_{3}-2\delta
s(Q^2)~].
\label{a8DIS}
\end{equation}
The value of $\Gamma_{1}^{p}(Q^2)$ at fixed $Q^2$ depends on the
extrapolation of $g_1$ used in the unmeasured $x$ region. Using
for $g_1$ in that region its perturbative QCD expression the E155
Collaboration obtained, from the analysis of the presently
available data, the following value for $\Gamma_{1}^{p}(Q^2)$ at
$Q^2=5~GeV^2$ \cite{E155}:
\begin{equation}
\Gamma_1^p(Q^2=5~GeV^2) = 0.118~\pm~0.004(stat)~\pm~ 0.007(syst).
\label{GammapE155}
\end{equation}
The values of $\Gamma_{1}^{p}(Q^2)$ reported by other
collaborations before the E155 data were published are very close
to that value (see, e.g., \cite{SMCexp}). Note that at very small
$x$ $g_1(x,Q^2)_{QCD}$ gives a negative contribution to
$\Gamma_1^p(Q^2)$. On the other hand, the E143 Collaboration has
reported \cite{E143} experimental values for $\Gamma_1^p(Q^2)$ at
different $Q^2$ using for $g_1$ in the unmeasured low $x$ region
Regge-type behaviour, and found at $Q^2=3~GeV^2$
\begin{equation}
\Gamma_1^p(Q^2=3~GeV^2) = 0.133~\pm~0.003(stat)~ \pm~ 0.009(syst).
\label{GammapE143}
\end{equation}
In this case the low $x$ contribution to $\Gamma_1^p$ is positive
and that is the main reason why the central value of $\Gamma_1^p$
in (\ref{GammapE143}) is significantly different from the central
value in (\ref{GammapE155}). Note that $\Gamma_1^p(Q^2)$ itself
varies very slowly with $Q^2$, so that it is not the change in
value of $Q^2$ that is responsible for the difference. Thus using
the values (\ref{GammapE155}) or (\ref{GammapE143}) for
$\Gamma_1^p$ in Eq. (\ref{a8DIS}), a non-negative strange quark
polarization, {\it i.e.}, $\delta s\geq 0$ requires either
\begin{equation}
a_{8}\leq 0.089~\pm~0.058
\label{a8E155}
\end{equation}
or
\begin{equation}
a_{8}\leq 0.197~\pm~0.068
\label{a8E143}
\end{equation}
respectively, in both cases significantly contradicting the
bounds in (\ref{rangea8}). Hence a non-negative value of $\delta
s$ would imply a total breaking of $SU(3)_f$ symmetry for the
strong interactions. We are thus forced to conclude that a
non-negative first moment of $(\Delta s + \Delta \bar s)(x)$ is
almost impossible.\\

HERMES has not published the numerical data on the actual
measured asymmetries, so, we can only speculate on possible
causes why their analysis favours  slightly positive values for
$(\Delta s+ \Delta \bar s)(x, Q^2)$ in the medium x range:

~~i) The HERMES analysis involves a Monte Carlo LUND model for
the {\it purity} functions, tuned to fit the measured
multiplicities. It is not clear to what extent this method is
compatible with the LO QCD approach involving products of parton
densities and genuine fragmentation functions.

~ii) Consistency aside, a recent study \cite{KLCh} showed that
the myth that fragmentation functions are very well known from
$e^{+}e^{-}\longrightarrow hX$ is unjustified and that they have
significant uncertainties. This is especially true of
$D_{s}^{\pi}(z, Q^2)$, which plays a crucial role, in QCD
analysis using directly the genuine fragmentation functions, in
determining $(\Delta s+ \Delta \bar s)(x, Q^2)$. From this point
of view it may be that the uncertainty attributed to $(\Delta s +
\Delta \bar s)(x, Q^2)$ in a standard LO QCD analysis will be
much larger than the uncertainty found by HERMES.

iii) It might be suggested that the mean transverse momentum of
the detected hadron in the HERMES experiment is too small
($<p_{T}>\simeq 0.5GeV$) to justify the parton model approach. We
do not think this is relevant since the fundamental scale which
determines the applicability of the parton model is $Q^{2}$ and
the value quoted above should be adequate. However, some care
must be exercised regarding higher twist and NLO effects. For
example, we have shown in the inclusive case that while higher
twist effects are negligible in the ratio $g_1/F_1$
\cite{LSS2001} they are important in $g_1$ itself
\cite{newpaper}. Something similar may happen in the
semi-inclusive case.

As mentioned, these are only speculations. Further progress in
understanding why HERMES finds marginally positive values for the
polarized strange quark densities must await the publication by
HERMES of their actual asymmetry data.
\\

{\large \bf Acknowledgments}
\vskip 4mm

One of us (D.S.) is grateful for the hospitality of the Theory
Division at CERN where this work has been completed. This
research was supported by a UK Royal Society Collaborative Grant
and by the Bulgarian National Science Foundation under Contract
Ph-1010.

\end{document}